\documentclass{PoS}
\usepackage{epsfig}

\title{Spectroscopic Analysis of subluminous B Stars in Binaries}

\ShortTitle{Spectroscopic Analysis of sdB Binaries}

\author{\speaker{S. Geier}$^1$, C. Karl$^1$, H. Edelmann$^2$, U. Heber$^1$ and R. Napiwotzki$^3$\\
\llap{$^1$}Dr.\,Remeis-Sternwarte\\
        Sternwartstra\ss e 7, 96049 Bamberg, Germany\\
\llap{$^2$}McDonald Observatory, University of Texas at Austin\\
        1 University Station, C1402, Austin, TX 78712-0259, USA\\
\llap{$^3$}Centre for Astrophysics Research, University of Hertfordshire\\ 
        College Lane, Hatfield AL10 9AB, UK\\
        
E-mail: \email{geier@sternwarte.uni-erlangen.de}, \email{karl@sternwarte.uni-erlangen.de}, 
        \email{edelmann@astro.as.utexas.edu}, \email{heber@sternwarte.uni-erlangen.de}, 
        \email{rn@star.herts.ac.uk}}

\abstract{ The masses of compact objects like white dwarfs, neutron stars and black holes are fundamental to astrophysics, but very difficult to measure. We present the results of an analysis of subluminous B (sdB) stars in close binary systems with unseen compact companions to derive their masses and clarify their nature.  Radial velocity curves were obtained from time resolved spectroscopy. The atmospheric parameters were determined in a quantitative spectral analysis. With high resolution spectra we were able to measure the projected rotational velocity of the stars with high accuracy. The assumption of orbital synchronization made it possible to constrain inclination angle and companion mass of the binaries. Five invisible companions have masses that are compatible with white dwarfs or late type main sequence stars. But four sdBs have very massive companions like heavy white dwarfs, neutron stars or even black holes. Such a high fraction of massive compact companions can not be explained with current models of binary evolution.}

\FullConference{International Symposium on Nuclear Astrophysics --- Nuclei in the Cosmos --- IX\\
		 June 25-30 2006\\
		 CERN, Geneva, Switzerland}

\begin{document}

\section{Introduction}

The mass of a star is it's most fundamental parameter. However, a direct measurement is possible in some binary stars only. Eclipsing, double lined systems are first choice. White dwarfs, neutron stars and stellar black holes are the aftermath of stellar evolution. In binaries such faint, compact objects are outshined by their bright companions and therefore their orbital motion cannot be measured. As a consequence only lower limits to the companion mass can be derived. With the analysis method shown here, these limitations can partly be overcome.\\
Subluminous B stars, which are also known as hot sudwarf stars (sdBs), show the same spectral characteristics as main sequence stars of spectral type B, but are much less luminous. They are considered to be helium core burning stars with very thin hydrogen envelopes and masses around $0.5\,M_{\rm \odot}$.
The formation of these objects is still puzzling. Different formation channels have been discussed. As it turned out, a large fraction of the sdB stars are members of short period binaries (Maxted et. al 2001 [10], Napiwotzki et al. 2004 [14]). For these systems common envelope ejection is the most probable formation channel (Han et al. 2002, 2003 [6,7]). In this scenario two main sequence stars of different masses evolve in a binary system. The heavier one will first reach the red giant phase and fill its Roche lobe. If the mass transfer to the companion is dynamically unstable, a common envelope is formed. Due to friction the two stellar cores loose orbital energy, which is deposited within the envelope and leads to a shortage of the binary period. Eventually the common envelope is ejected and a close binary system is formed, which contains a helium core burning sdB and a main sequence companion. If this star reaches the red giant branch, another common envelope phase is possible and can lead to a close binary with a white dwarf companion and an sdB. All known companions of sdBs in such systems are white dwarfs or late type main sequence stars. If massive stars are involved, the primary may evolve into a neutron star (NS) or a black hole (BH) rather than a white dwarf. Since massive stars are very rare, only few sdB+NS or sdB+BH systems are expected to be found.\\
Since the spectra of the program stars are single-lined,
 they reveal no information about the orbital motion of the
 sdBs' companions, and 
 thus only their mass functions can be calculated.

 \begin{equation}
 \label{equation-mass-function}
 f_{\rm m} = \frac{M_{\rm comp}^3 \sin^3i}{(M_{\rm comp} +
   M_{\rm sdB})^2} = \frac{P K^3}{2 \pi G} .
 \end{equation}

Although the RV semi-amplitude $K$ and the period $P$ are determined
 by the RV curve, $M_{\rm sdB}$, $M_{\rm comp}$ and $\sin{i}$ remain
 free parameters.
Binary population synthesis models (Han et al. 2002, 2003 [6,7]) indicate a possible mass range of $M_{\rm sdB}$\,=\,0.30$-$0.48\,M$_{\rm \odot}$ for sdBs in binaries, which underwent the common envelope ejection channel. The mass distribution shows a sharp peak at about $0.46\,M_{\rm \odot}$. \\
For close binary systems, the components' stellar rotational velocities are considered to be tidally locked to their orbital motions (Zahn 1977 [17], Tassoul \& Tassoul 1992 [16]), which means that the orbital period of the system equals the rotational period of the companions. If the companions are synchronized in this way the rotational velocity $v_{\rm rot}$ can be calculated.

\begin{equation}
v_{\rm rot} = \frac{2 \pi R_{\rm sdB}}{P} .
\end{equation}

The stellar radius $R$ is given by the mass radius relation.

\begin{equation}
R = \sqrt{\frac{M_{\rm sdB}G}{g}}
\end{equation}

The measurement of the projected rotational velocities 
 $v_{\rm rot}\,\sin\,i$ 
therefore allows to constrain the systems' inclination angles $i$.
With $M_{\rm sdB}$ as free parameter the mass function can be solved and the inclination angle as well as the companion mass can be derived. Because of $\sin{i} \leq 1$ a lower limit for the sdB mass is given by

\begin{equation}
M_{\rm sdB} \geq \frac{v_{\rm rotsini}^{2} P^{2}g}{4 \pi^{2}G}
\end{equation}

To constrain the system parameters in this way it is necessary to measure $K$, $P$, $\log{g}$ and $v_{\rm rot}\sin{i}$ with high accuracy. 

\begin{figure}[t!]
 \begin{center}
 \vbox{
 \centerline{\psfig{figure=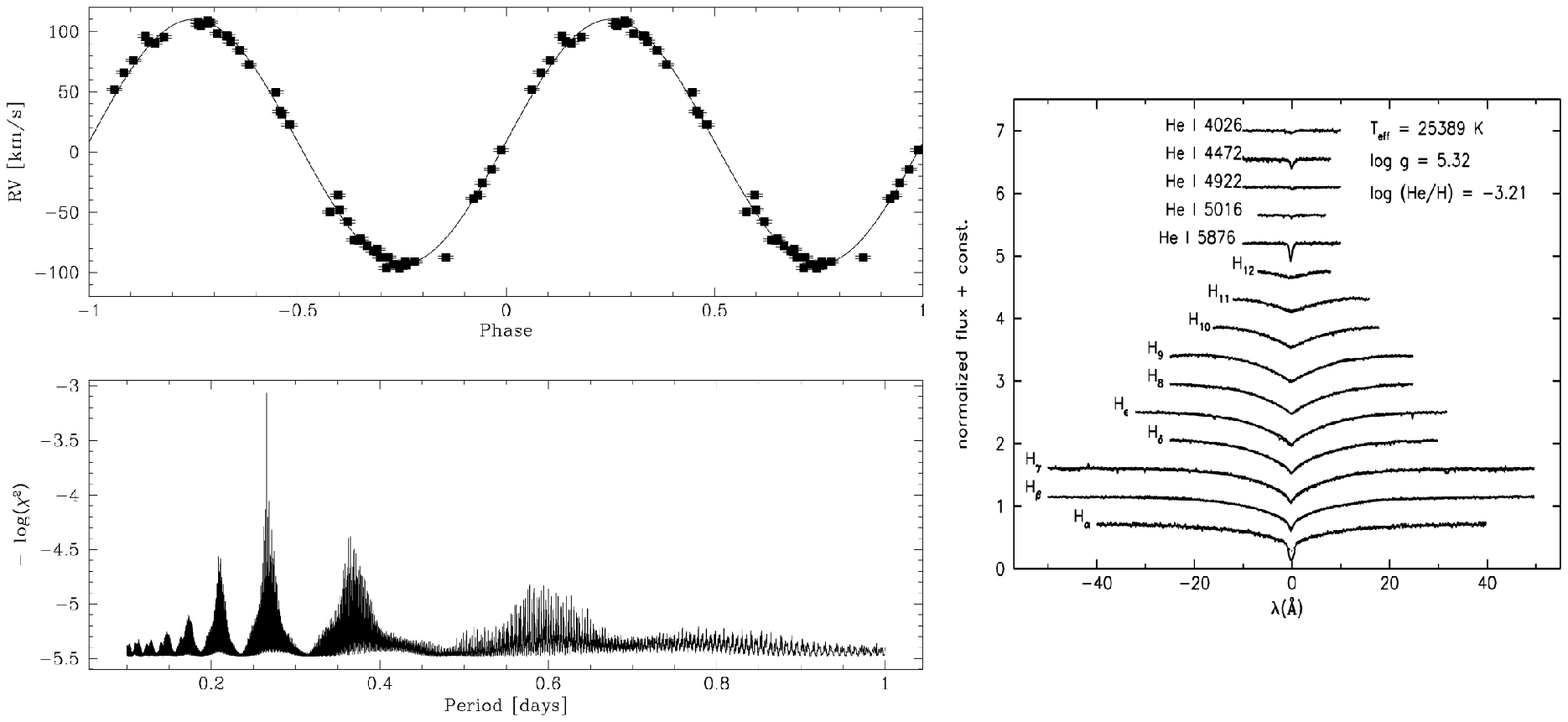,width=150truemm,angle=0}}
 {\parbox{125mm}{
 {\bf Figure~1.}{ 
 Sample model fit for an sdB star (HE\,0532$-$4503) based on
 47 UVES spectra (right panel).
 Sample best fit RV curve and power spectrum for the visible sdB star
 in the HE\,0532$-$4503 system (left panel).
 Upper left panel: Measured radial velocities as a
 function of orbital phase and fitted sine curve.
 Lower left panel: Power spectrum.
 }}}
 }
 \end{center}
 \label{RVfit}
 \end{figure}

\begin{figure}[t!]
 \begin{center}
 \vbox{
 \centerline{\psfig{figure=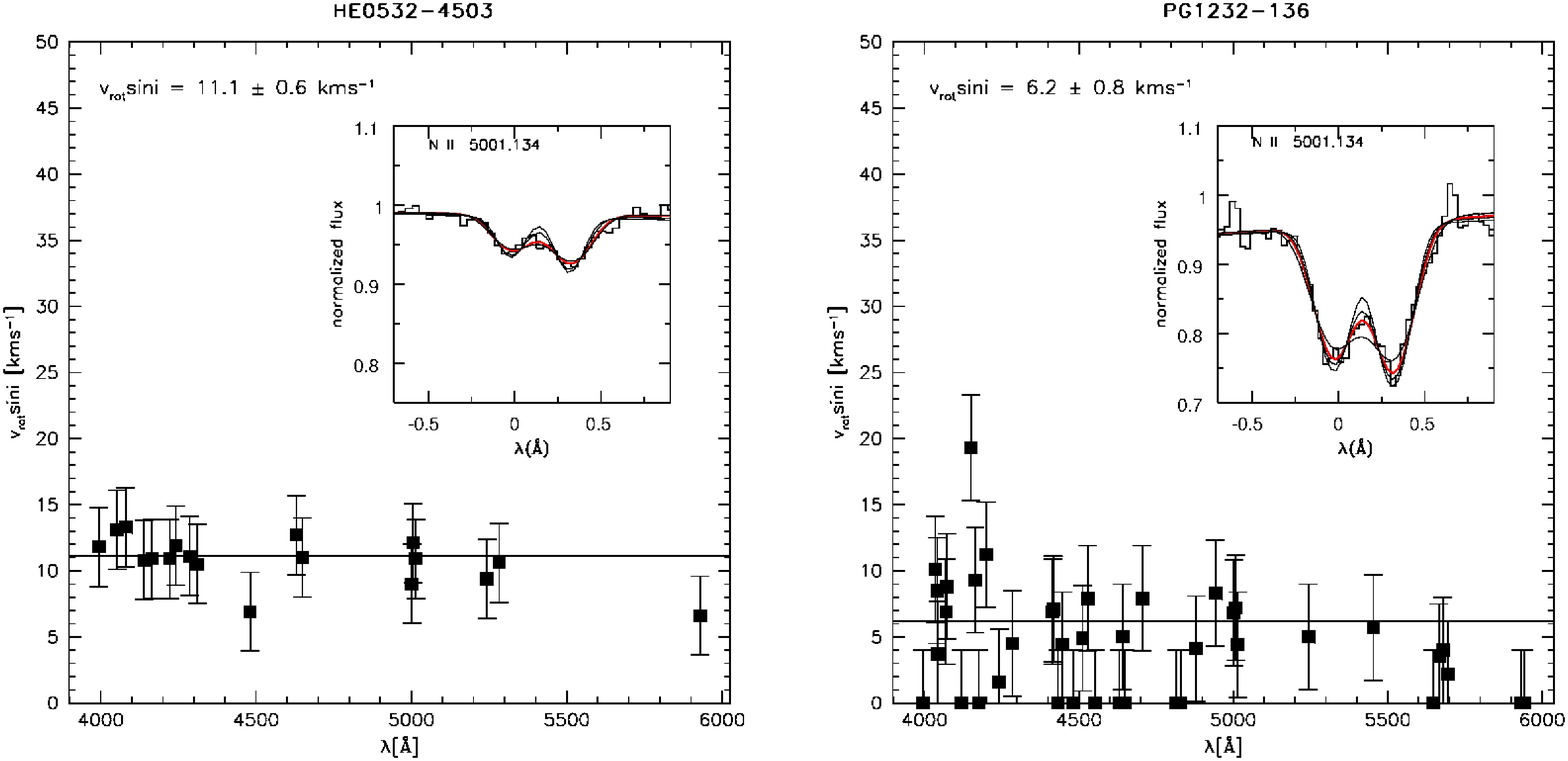,width=150truemm,angle=0}}
 {\parbox{125mm}{
 {\bf Figure~2.}{ 
 Projected rotational velocity as a function of wavelength for the sample stars HE\,0532$-$4503 (left panel) and PG\,1232$-$136 (right panel). The filled squares are measurements of single metal lines. The mean value  
 is plotted as black horizontal line. The small boxes show a sample fit,  
 where the red line marks the best solution and the narrow black lines fixed rotational broadenings of $0,\,5$ and $10\,{\rm kms^{-1}}$
 }}}
 }
 \end{center}
 \label{vrotsini}
 \end{figure}

\section{Observations and Radial Velocity Curves}
Ten stars were observed at least twice with the high resolution spectrograph UVES at the ESO\,VLT.
Additional observations were made at the ESO\,NTT (equipped with EMMI), the Calar Alto Observatory 3.5\,m telescope
 (TWIN) and the 4\,m WHT (ISIS) at La Palma (Napiwotzki et al. 2003 [11]).
Two of the stars (PG\,1232$-$136, TONS\,183) were observed with the high resolution FEROS instrument at the 2.2\,m ESO telescope at La Silla (Edelmann et al. 2005 [5]).
The data was reduced with different software packages specially developed for every instrument.\\
Radial velocities of the individual observations were determined
 by calculating the
 shifts of the measured wavelengths relative to their laboratory values. 
We focussed on
 all available He\,{\sc i} lines, and on
 the observed H${\alpha}$ line profile because of its
 sharp and well-defined non-LTE line core.
We performed a simultaneous fit of a set of mathematical functions to the
 observed line profiles using the ESO MIDAS package.
A linear function was used to reproduce the overall spectral trend,
 and a Gaussian for the innermost line core.
In order to fit the H${\alpha}$ profile we used an additional
  Lorentzian to model the broad line wings.
The central wavelength of the Lorentzian was fixed to that of the Gaussian
 for physical reasons.
The period search was carried out by means of a periodogram
 based on the 'Singular Value Decomposition' method.
For a large range of periods the best fitting sine-shaped RV curve 
was computed (see Napiwotzki et
 al.~2001 [12]).
The difference between the observed radial velocities and the
 best fitting theoretical RV curve
 for each phase set was evaluated in terms of the logarithm of 
 the sum of the squared
 residuals ($\chi^2$) as a function of period.
This method finally results in the data-set's power spectrum which
 allows to determine
 the most probable period of variability 
 (see Lorenz et al.~1998 [9]).\\
From the best fit RV curve corresponding to the most probable period, the ephemeris, the
 system's velocity and the semi-amplitude were derived.
As an example, 
 Fig.~1 (left panel) displays the resulting power spectrum and 
 best-fit sine curve for HE~0532$-$4503.
The orbital parameters of TONS\,183 and PG\,1232$-$136 were taken from Edelmann et al. (2005) [5]. An orbital solution for HE\,1047$-$0436 was published by Napiwotzki et al. (2001) [12].
The orbital periods ($P$) and the radial velocity semi-amplitudes ($K$) are given in Tab.~1.

\section{Quantitative Spectral Analysis}
Prior to quantitative spectral analysis the spectra were corrected for the 
 measured RV and coadded in order to increase
 the S/N ratio.
Effective temperatures ($T_{\rm eff}$), surface gravities ($\log g$) and
 helium abundances ($\log\,[n_{\rm He}/n_{\rm H}]$) were
 determined by fitting simultaneously each observed hydrogen and helium line
 with a grid of metal-line blanketed LTE model spectra.
The procedure used is described in detail in Napiwotzki et al. (1999) [13] and Lisker et al. (2005) [8].
Because of its sensitivity to non-LTE effects, the H$\alpha$ line was
 excluded from this analysis.
The atmospheric parameters of TONS\,183 and PG\,1232 $-$ 136 were taken from Edelmann et al. (2005) [5].
Results are displayed in Tab.~1, and a sample
 fit is shown in Fig.~1 (right panel).

\section{Projected Rotational Velocity}

In order to derive $v_{\rm rot}\,\sin\,i$,
 we compared the observed spectra 
 with rotationally broadened, synthetic line profiles.
The latter ones were computed for the stellar
 parameters given in Tab.~1 using the LINFOR program.

 Since sharp metal lines
 are much more sensitive to rotational broadening
 than Balmer or helium lines, all visible metal lines were included. A simultaneous fit of elemental abundance and projected rotational velocity was performed separately for every identified line using the FITSB2 routine (Napiwotzki et al. 2004b [15]). Mean value and standard deviation were calculated from all measurements (see Fig.~2). Numerical simulations were carried out to identify potential sources of errors, the sensitivity limit and accuracy of the method. As the lower limit we derived $v_{\rm rot}\sin{i} > 5\, {\rm kms^{-1}}$.\\
All other possible sources of systematic errors turned out to be negligible compared to the statistical variations. No significant microturbulence could be measured, which is consistent with the analysis of Edelmann (2003) [4]. Unconsidered effects would in any case cause an extra broadening of the lines. This fact is important for the interpretation of the results and will be discussed in detail later.\\

\begin{figure}[t!]
 \begin{center}
 \vbox{
 \centerline{\psfig{figure=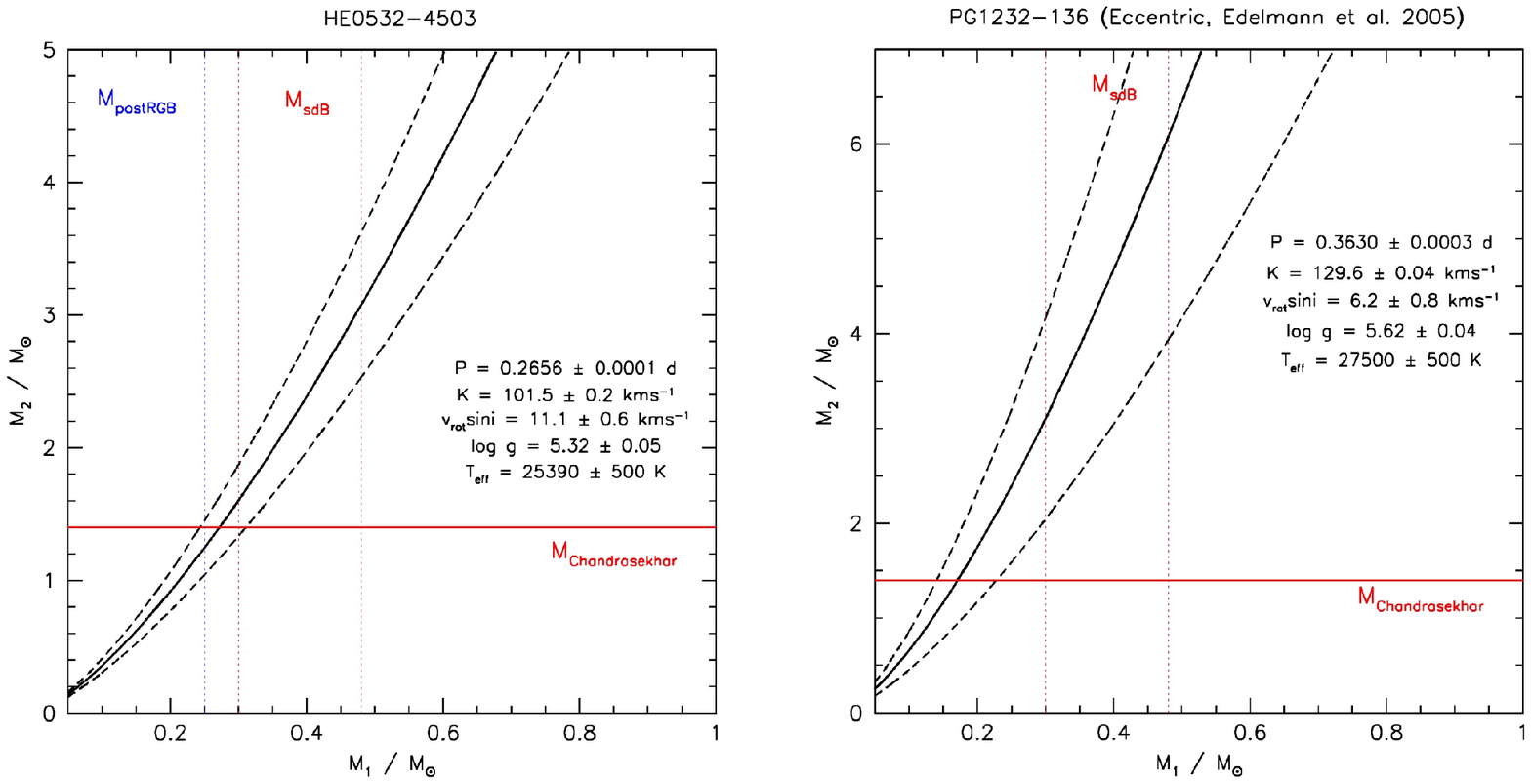,width=140truemm,angle=0}}
 {\parbox{125mm}{
 {\bf Figure~3.}{ 
 Companion mass as a function of primary (sdB) mass of the binaries HE\,0532$-$4503 (left panel) and PG\,1232$-$136 (right panel). The red horizontal line marks the Chandrasekhar limit. The red dotted vertical lines mark the theoretical sdB mass range for the common envelope ejection channel (Han et al. 2002 [6]). The dotted blue line in the left panel marks the primary mass in case of post-RGB evolution (Driebe et al. 1998 [3]).
 }}}
 }
 \end{center}
 \label{lock1}
 \end{figure}

\begin{figure}[t!]
 \begin{center}
 \vbox{
 \centerline{\psfig{figure=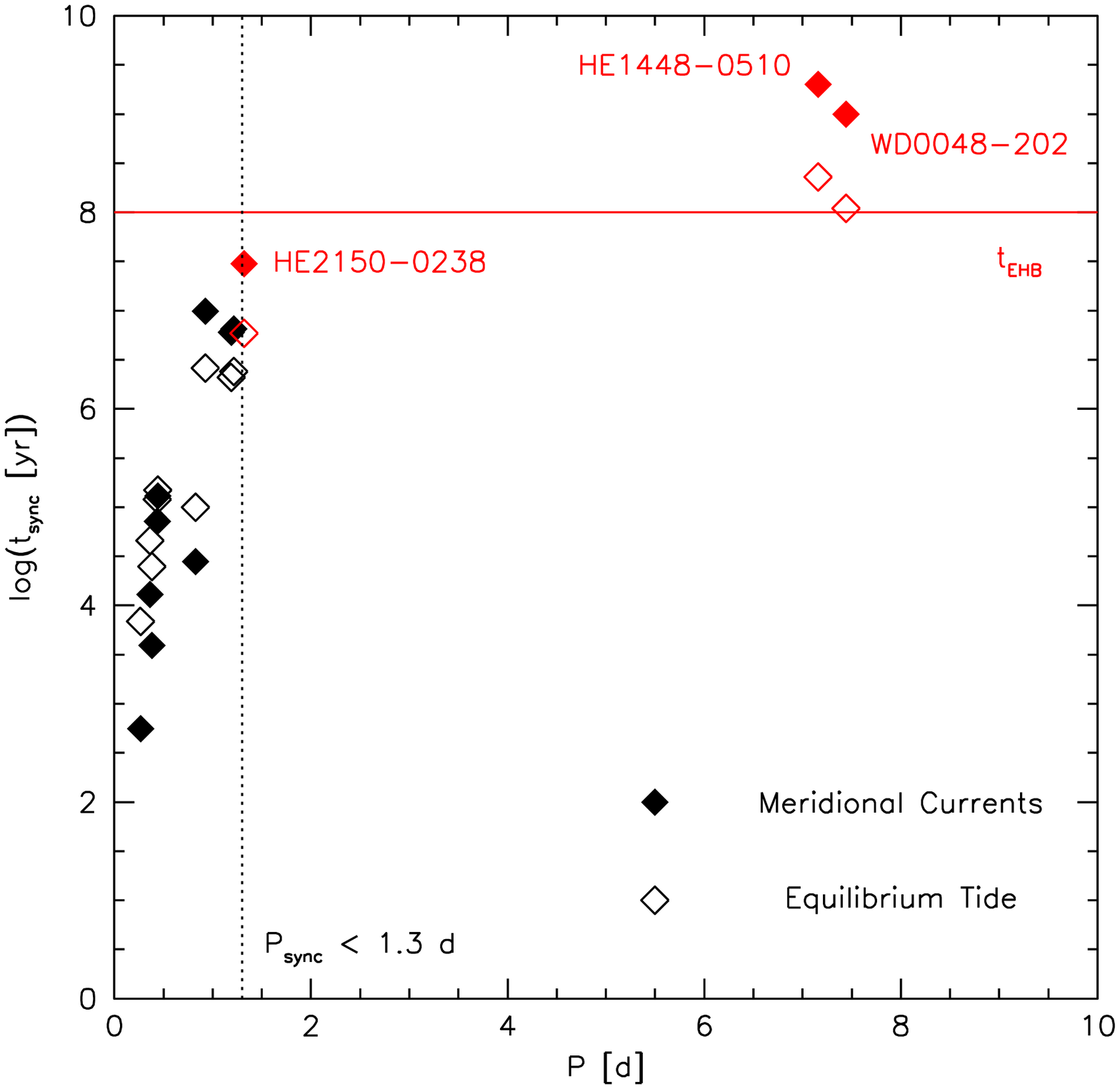,width=75truemm,angle=0}}
 {\parbox{125mm}{
 {\bf Figure~4.}{ 
 Synchronization time over orbital period of the 12 analysed binaries. The filled diamonds mark synchronization caused by meridional currents in radiative envelopes (Tassoul \% Tassoul 1992) [16], the blank ones synchronization through tidal friction of the equilibrium tide for stars with radiative envelope and convective core (Zahn 1977) [17]. The horizontal line marks the lifetime on the Extreme Horizontal Branch (EHB). The dotted vertical line denotes the limiting orbital period for tidal synchronization derived from our analysis.
 }}}
 }
 \end{center}
 \label{tidal}
 \end{figure}

\section{Nature of the unseen Companions}

To constrain the companion masses and inclination angles we adopted the mass range for sdBs from Han et al. (2002) [6]. There are no spectral signatures of companions visible. Main sequence stars with masses higher than $0.45\,M_{\rm \odot}$ could therefore be excluded because of their high luminosities in comparison to the sdB stars. The possible companion masses can be seen in Tab.~2. Four of the analysed systems have companion masses, which are compatible with either white dwarfs (WD) or late main sequence stars (late MS). The companion of WD\,0107$-$0342 is a white dwarf.\\
The very similar binaries HE\,0929$-$0424 and TONS\,183 have a wide companion mass range. For very low mass sdBs the companion could be a heavy white dwarf. But taking into account, that the theoretical sdB mass distribution has a sharp peak at $0.46\,M_{\rm \odot}$, which makes this the most probable sdB mass, the companion masses exceed the Chandrasekhar limit. There are only two kinds of objects known with such high masses and such low luminosities - neutron stars (NS) and stellar black holes (BH). The two systems HE\,0532$-$4503 and PG\,1232$-$136 have even higher companion masses, which exceed the Chandrasekhar limit in any case.\\
The three binaries HE\,1448$-$0510, HE\,2150$-$0238 and WD\,0048$-$202 could not be solved with the described method. The minimum sdB masses exceeded $1.3\,M_{\rm \odot}$ and were not consistent with the theoretical mass range.\\
Binaries hosting a neutron star or a black hole are a very rare object class. From about 50 analysed sdB binaries in our samples, we found four candidate systems. This fraction of 8 \% is much too high to be compatible with any binary evolution model known so far (Podsiadlowski priv. comm.). For this reason the results themselves have to be discussed again.\\
Our method rests on the assumption of orbital synchronization. The question, which mechanism is most responsible for this effect, is not yet settled. Theoretical timescales for synchronization are given by Zahn (e.g. 1977) [17] and Tassoul \& Tassoul (1992) [16], but unfortunately they are not consistent in any case. Many eclipsing binaries in close orbits were observed to be synchronized (e.g. Claret et al. 1996, 1997 [1,2]).\\  
The approximation formulas for the synchronization time of Zahn (1977) [17] and Tassoul \& Tassoul (1992) [16] for stars with radiative envelopes and convective cores were used to compare them with the lifetime on the Extreme Horizontal Branch (EHB) $t_{\rm EHB} \approx 10^{8}\,{\rm yr}$ (Fig. 4 right panel). The red diamonds denote the three systems, which could not be solved consistently. These three binaries have the longest orbital periods and synchronization times near or above the EHB lifetime. But as long as the question of tidal synchronization is not settled, all timescales have to be taken with caution. Detailed calculations, which take the internal structure of sdBs into account, are not available and urgently needed.\\
From these considerations we draw the conclusion, that the assumption of tidally locked rotation is reasonable in case of close sdB binaries for orbital period shorter than $1.3\,d$.\\
Finally the reliability of our method has to be discussed again. Only the high precision of the $v_{\rm rot}\sin{i}$ measurements made it possible to constrain the parameters reasonably. The projected rotational velocities were low and very close to the detection limit. As described above we tried to quantify all possible systematic effects and the overall results were very consistent. Only slight changes in $v_{\rm rot}\sin{i}$ would lead to inconsistent solutions, because the method is very sensible to this parameter. But even if there would be unaccounted systematic effects (e.g. short period pulsations), they would always cause an extra broadening of the lines. The measured broadening is then due to rotation and the unaccounted effects, which means the measured rotational broadening is higher than the real one. All systematic effects lead to lower $v_{\rm rot}\sin{i}$. But the unexpectedly high masses of the companions are caused by the unexpectedly low measured projected rotational velocities as can be seen from the equations in the introduction. Unaccounted systematic effects would therefore lead to even higher companion masses.\\

\begin{center}
\vbox{

\begin{tabular}{l|cccccc}

System & $T_{\rm eff}$ & $\log{g}$ & $P$ & $K$ &  $v_{\rm rot}\,\sin\,i$ \\
       & [K] &  & [d] & [${\rm km\,s^{-1}}$] & [${\rm km\,s^{-1}}$] \\ 
\hline
HE\,0532$-$4503 & 25390 & 5.32 & 0.26560 $\pm$ 0.00010 & 101.5 $\pm$ 0.2 & 11.1 $\pm$ 0.6 \\
PG\,1232$-$136 & 27500 & 5.62 & 0.36300 $\pm$ 0.00030 & 129.6 $\pm$ 0.04 & 6.2 $\pm$ 0.8 \\
WD\,0107$-$342$\dag$ & 24300 & 5.32 & 0.38000 $\pm$ 0.00050 & 135.0 $\pm$ 1.0 & 20.4 $\pm$ 0.9 \\
HE\,0929$-$0424 & 29470 & 5.71 & 0.44000 $\pm$ 0.00020 & 114.3 $\pm$ 1.4 & 7.1 $\pm$ 1.0 \\
HE\,0230$-$4323 & 31100 & 5.60 & 0.44300 $\pm$ 0.00050 & 64.1 $\pm$ 1.5 & 12.7 $\pm$ 0.7 \\
TONS\,183 & 26100 & 5.20 & 0.82770 $\pm$ 0.00020 & 84.8 $\pm$ 1.0 & 6.7 $\pm$ 0.7 \\
HE\,2135$-$3749 & 30000 & 5.84 & 0.92400 $\pm$ 0.00030 & 90.5 $\pm$ 0.6 & 6.9 $\pm$ 0.5 \\
HE\,1421$-$1206 & 29570 & 5.55 & 1.18800 $\pm$ 0.00100 & 55.5 $\pm$ 2.0 & 6.7 $\pm$ 1.1 \\
HE\,1047$-$0436 & 30242 & 5.66 & 1.21325 $\pm$ 0.00001 & 94.0 $\pm$ 3.0 & 6.2 $\pm$ 0.6 \\
HE\,2150$-$0238 & 30200 & 5.83 & 1.32090 $\pm$ 0.00500 & 96.3 $\pm$ 1.4 & 8.3 $\pm$ 1.3 \\
HE\,1448$-$0510 & 34690 & 5.59 & 7.15880 $\pm$ 0.01300 & 53.7 $\pm$ 1.1 & 6.7 $\pm$ 2.5 \\
WD\,0048$-$202  & 29960 & 5.50 & 7.44360 $\pm$ 0.01500 & 47.9 $\pm$ 0.4 & 7.2 $\pm$ 1.3 \\
\hline
\\
\end{tabular}
{\parbox{125mm}{
{\bf Table 1.}{
Stellar parameters:
Effective temperatures, surface gravities, orbital periods, radial velocity semi-amplitudes and projected rotational velocities
of the visible components.
Typical error margins for  $T_{\rm eff}$ and $\log g$ are
500\,K and 0.05\,dex respectively.\\
$\dag$ Orbital parameters of this system are still preliminary
}}}\\
}
\end{center}

\begin{center}
\vbox{
\begin{tabular}{l|cccccc}

System &  $i$ & $v_{\rm rot}$ & $M_{\rm comp}$ & Companion\\
       & [deg] & [${\rm km\,s^{-1}}$] & [$M_{\rm \odot}$] \\ 
\hline
HE\,0532$-$4503 & 13 $-$ 17 & 47 & 1.40 $-$ 3.60 & \bf NS/BH\\
PG\,1232$-$136 & 14 $-$ 19 & 25 & 2.00 $-$ 7.00 & \bf NS/BH \\
WD\,0107$-$342$\dag$ & 17 $-$ 22 & 33 & 0.45 $-$ 0.95 & WD \\
HE\,0929$-$0424 & 23 $-$ 29 & 18 & 0.60 $-$ 2.40 & \bf WD/NS/BH \\
HE\,0230$-$4323 & 38 $-$ 50 & 21 & 0.18 $-$ 0.35 & WD/late MS\\
TONS\,183 & 22 $-$ 29 & 18 & 0.60 $-$ 2.40 & \bf WD/NS/BH \\
HE\,2135$-$3749 & 66 $-$ 90 & 8 & 0.35 $-$ 0.45 & WD/late MS\\
HE\,1421$-$1206 & 56 $-$ 90 & 8 & 0.15 $-$ 0.30 & WD/late MS\\
HE\,1047$-$0436 & 62 $-$ 90 & 7 & 0.35 $-$ 0.60 & WD/late MS\\
HE\,2150$-$0238 & -- & -- & -- & not solved \\
HE\,1448$-$0510 & -- & -- & -- & not solved \\
WD\,0048$-$202  & -- & -- & -- & not solved \\
\hline\\
\end{tabular}
{\parbox{125mm}{
{\bf Table 2.}{
Inclination angles, rotational velocities, companion masses and possible nature of the unseen companions.\\
$\dag$ Derived parameters of this system are still preliminary
}}}\\
}
\end{center}

\pagebreak

\section{Conclusion}

Out of 12 analysed sdB binaries, five have companion masses compatible with white dwarfs or late type main sequence stars. These systems are in full agreement with binary evolution simulations. Four binaries have surprisingly high companion masses, which leads to the conclusion, that the companions have to be very heavy white dwarfs or even neutron stars or black holes. This high fraction cannot be explained with current evolutionary calculation. More observations of close sdB binaries and candidate systems with high and low resolution spectrographs are needed to enhance the size of our samples.\\
The presence of such a high fraction of heavy binaries in our samples opens several questions. Is the formation and evolution of sdB stars somewhat linked to that of heavy compact objects like neutron stars or black holes? Can sdB stars be used as tracers to find more of these exotic objects? Is there a hidden population of these objects, which is not visible through strong X-ray emission?\\

\vskip1mm

\end{document}